\documentstyle[aps,prl,multicol,epsf]{revtex}
\begin{document}
\widetext
\title{\bf Hall coefficient and ARPES in systems with strong
pair fluctuations}

\author{Alfonso Romano$^{(a)}$ and Julius Ranninger$^{(b)}$}

\address{$^{(a)}$ Dipartimento di Scienze Fisiche "E.R. Caianiello",
Universit\`a di Salerno, I-84081 Baronissi (Salerno), Italy\\
Unit\`a I.N.F.M. di Salerno}

\address{$^{(b)}$ Centre de
Recherches sur les Tr\`es Basses Temp\'eratures,
Laboratoire
Associ\'e \`a l'Universit\'e Joseph Fourier,
\\ Centre National de la
Recherche Scientifique, BP 166, 38042, Grenoble
C\'edex 9, France}

\date{\today}
\maketitle
\draft
\begin{abstract}
We examine the  normal-state temperature and doping
dependence of the Hall coefficient in the context of a
pair-fluctuation scenario, based on a model where itinerant electrons
are hybridized  with localized electron pairs via a charge
exchange term. We show that an anomalous behavior of the Hall effect,
qualitatively similar to that observed in high-$T_c$ superconductors,
can be attributed to the non-Fermi liquid properties
of the single-particle spectral function which exhibits pseudogap
features. Our calculations are based on a dynamical mean-field
procedure which relates the transport coefficients to the
single-particle spectral function in an exact way.

\end{abstract}


\begin{multicols}{2}

\narrowtext
\section{INTRODUCTION}
The anomalous temperature and doping behavior of
transport coefficients\cite{experiments} such as the dc
resistivity $\rho$, the Hall coefficient $R_H$, the
magnetoresistance $\Delta \rho$ and the Hall
angle $cot \, \theta_H = \rho/R_H$ in the normal state of
high-$T_c$ cuprate superconductors (HTS) are amongst
the clearest indications for non-Fermi liquid behavior in these
materials.  Quite generally $R_H$ shows a rapid decrease with
increasing temperature and in certain overdoped samples can change
from positive
to small negative values. This, together with the Hall angle
showing an unexpected quadratic temperature variation
in the underdoped regime, has led to speculations
about different transport mechanisms in presence and in
absence of a magnetic field\cite{Anderson-91}, based on the
phenomenological introduction of two different scattering rates -
perpendicular and parallel to the Fermi surface. Such a
description may hold in non-Fermi liquids, provided they
retain sufficient Fermi liquid properties so that the concept
of scattering rates and a description in terms of a Boltzmann equation
is still meaningful. When this is not guaranteed, the interpretation
of the anomalous transport properties has to be sought in the
single-particle spectral features themselves, strongly affected by
many-body effects.  The Mott-Hubbard
scenario of a strongly correlated system\cite{Pruschke-95,Kotliar-98}
as well as  the $t-J$ spin
correlation picture\cite{Shastry-93}, which are  believed
to capture certain features of the underdoped HTS, have been
investigated in this connection
and show qualitative similarities with the experimental results.

We shall in this Letter address the question of the anomalous
magneto-transport processes in terms of a scenario where the
Fermi liquid properties get destroyed in the normal state  due
to dynamical electron  pair
formation  showing up in the appearence of a pseudogap. The
decrease of $R_H$ with increasing temperature is then expected to be
related to thermal excitations across such a pseudogap feature in the
single-particle spectra. The present study aims to
examine the underdoped and the overdoped systems on equal footing.

It is presently a matter of debate whether the onset of the pseudogap
at $T^*$  is a precursor of the superconducting state or not. If not,
pair fluctuations are a necessary
but not a sufficient condition for the system to become
superconducting upon lowering the temperature. The present
experimental situation seems to suggest that the opening of the
pseudogap is
not triggered by superconducting precursor effects. If it were,
then superconducting correlations should show up in the entire
temperature regime of the pseudogap phase, giving rise to an
enhancement of the dc conductivity, the
diamagnetism etc. into a Meissner state.
Such features are indeed observed, but only in a very reduced
fraction of the pseudogap temperature regime relatively close
to $T_c$, where such pair fluctuations exist on a time scale of
the order of $T_c$. Recent experiments on the frequency dependence of
the Meissner screening\cite{Corson-98} seem also to confirm that.

Presently there are two phenomenological models, the negative-U
Hubbard model and the Boson-Fermion model (BFM), which are studied
in connection with those experimental findings. In the negative-U
Hubbard model the HTS physics is treated as a cross-over phenomenon
between a BCS state in the overdoped regime and something close to a
Bose-Einstein condensation (BEC) of preformed pairs in the underdoped
regime\cite{Randeria-98}. Going from one extreme doping situation to
the other one is monitored by a variation of the attractive interaction U
between the electrons. In the BFM the HTS physics is described by
a mechanism by which pairing in an electronic subsystem is induced
by resonant scattering of the electrons in and out of localized
tightly bound pair electron (bosonic) states. Doping is monitored by
changing the position of the bosonic energy level which alters the relative
occupation between the electrons and the bosons. The bosonic states
in the BFM have their counterpart in the negative-U Hubbard system
in form of two-particle resonant states above the chemical
potential.

The BFM was originally introduced\cite{Ranninger-85} and studied
in some detail in the past as regards the single-particle spectral
properties of the electrons\cite{Ranninger-95}. The
qualitative differences which exist between the underdoped and the
overdoped regime warrant to examine whether and how
these differences can possibly be held responsible for the unusual
temperature and doping  dependence of certain magneto-transport properties
seen in the HTS. Such a study must be done in a way which relates the
above mentioned transport coefficients to
the single-particle properties in a rigorous fashion. In the limit
of infinite dimensions this can be achieved  within the dynamical
mean-field theory approach which we shall adopt here.

The hamiltonian for the BFM is given by
\begin{eqnarray}
H & = & \varepsilon_0\sum_{i,\sigma}c^{\dagger}_{i\sigma}c_{i\sigma}
-t\sum_{\langle i j\rangle,\sigma}c^{\dagger}_{i\sigma}c_{j\sigma}
\nonumber \\
& & + \; E_0 \sum_i b^{\dagger}_i b_i
+ g \sum_i [b^{\dagger}_i c_{i\downarrow} c_{i\uparrow}
+ c^{\dagger}_{i\uparrow} c^{\dagger}_{i\downarrow} b_i] \quad .
\label{eq2}
\end{eqnarray}
\noindent
$c_{i\sigma}^{(\dagger)}$ denote  annihilation (creation) operators
for electrons with spin $\sigma$ at some effective sites $i$
(involving molecular units rather than individual atoms) and
$b_i^{(\dagger)}$ denote hard-core bosonic operators describing
tightly bound electron pairs.  $t$, $D$, $\Delta_B$ and  $g$ denote
respectively the bare hopping integral for the electrons, the bare
electronic half bandwidth, the boson energy level and the
boson-fermion pair-exchange coupling constant. Furthermore we put
$\varepsilon_0=D-\mu$ and $E_0=\Delta_B-2\mu$ and assume the
chemical potential $\mu$ to be common to fermions and bosons
(up to a factor 2 for the bosons) in order to guarantee charge
conservation. These parameters are fixed in such a way that at
temperatures large compared to the interaction $g$ the number of
fermions
$n_F=\sum_{i,\sigma }\langle c_{i \sigma}^{\dagger}c_{i \sigma} \rangle$
lies in the interval $[1, 0.75]$ which covers the typical experimental
doping regime. We furthermore choose $n=n_F+2n_B$
($n_B=\frac{1}{N}\sum_{i}\langle b_i^{\dagger} b_i \rangle$
denoting the number of bosonic electron pairs) to lie in the
interval $[1,2]$ in order to account for the appearence of a
pseudogap phase. For the present analysis we choose $n=1.5$
as a representative value, having verified that in the regime
$n=[1.2,1.8]$ the various physical quantities which we
shall discuss here exhibit qualitatively the same behavior.
Finally, in order to have a $T^*$ of the order of a few hundred
degrees K we choose $g=0.2$ (all energies are in units of the bare
electronic
bandwidth $2D$). From our previous studies we know that the
anomalous electronic properties in the normal state are largely
determined by those on an atomic or molecular level\cite{Domanski-98}.
The dynamical mean-field approach permits to
properly deal with the local electronic structure which is renormalized
by the itinerancy of the electrons via a Weiss
mean-field\cite{Georges-96}.

\section{ON THE ORIGIN OF THE PSEUDOGAP}
In order to highlight the physics leading up to the pseudogap
phenomenon, let us first consider the above Hamiltonian in the
atomic limit i.e., $t_{ij} = 0$\cite{Domanski-98}.
In that case we derive the following set of eigenstates $| n \rangle $
and eigenvalues $E_{n}$:
\begin{eqnarray}
| 1 \rangle \; &=& \; | 0 \rangle,\qquad\qquad\qquad\quad E_{1}
\; = \; 0 \qquad\qquad\qquad\qquad\nonumber \\
| 2 \rangle \; &=& \; | \uparrow \rangle \;,\qquad\qquad\qquad\;\;
E_{2} \; = \; \varepsilon_{0}
\nonumber \\
| 3 \rangle \; &=& \; | \downarrow \rangle \;,\qquad\qquad\qquad\;\;
E_{3} \; = \; \varepsilon_{0}
\nonumber \\
| 4 \rangle \; &=& \; u | \uparrow \downarrow \rangle \;
- \; v | \bullet \rangle,\qquad\; E_{4} \; = \; \varepsilon_{0} +
E_{0}/2 - \gamma
\nonumber \\
| 5 \rangle \; &=& \; v | \uparrow \downarrow \rangle \;
+ \; u | \bullet \rangle,\qquad\; E_{5} \; = \; \varepsilon_{0} +
E_{0}/2 + \gamma
\nonumber \\
| 6 \rangle \; &=& \; | \uparrow \bullet \rangle,\qquad\qquad\qquad
E_{6} \; = \;
\varepsilon_{0} + E_{0}
\nonumber \\
| 7 \rangle \; &=& \; | \downarrow \bullet \rangle,\qquad\qquad\qquad
E_{7} \; = \;
\varepsilon_{0} + E_{0}
\nonumber \\
| 8 \rangle \; &=& \; | \uparrow \downarrow \bullet \rangle,\qquad
\qquad\quad \;\; E_{8} \; = \;
2 \varepsilon_{0} + E_{0}
\end{eqnarray}
The presence on a given site of an electron with spin up and
down, respectively, is denoted by corresponding arrows and the presence
of a Boson is denoted by a dot. Moreover we have
\begin{eqnarray}
u^{2} \; &=& \; \frac{1}{2} \left[ \; 1 \; - \; \frac{\varepsilon_{0} -
E_{0}/2}{\gamma} \; \right],\;\;
v^{2} \; = \; \frac{1}{2} \left[ \; 1 \; + \; \frac{\varepsilon_{0} -
E_{0}/2}{
\gamma} \; \right] \nonumber \\
\gamma \; &=& \; \left[ \; (\varepsilon_{0} - E_{0}/2)^{2} \; + \; g^{2} \;
\right]^{1/2}, \qquad\quad uv \; = \; \frac{g}{2 \; \gamma}
\end{eqnarray}
The corresponding atomic Green's function takes the form
\begin{eqnarray}
G_0(i\omega_n) \; =  \; - \int_0^{\beta} d\tau e^{i\omega_n\tau}
< T \; [ \; c_{\uparrow}(\tau) \;
c_{\uparrow}^{\dagger} \; ] > \; = \; \nonumber \\
\frac{Z_{f}}{i\omega_n - \varepsilon_{0}} \; + \;
\left( 1 - Z_{f} \right)
\left[ {u^2 \over i \omega_n -E_4 + \varepsilon_0}  \; + \;
{v^2 \over i \omega_n - E_5 + \varepsilon_0} \right]
\end{eqnarray}
where the spectral weight of the
non-bonding single-particle excitations is given by
\begin{equation}
Z_{f} \; = \; \frac{1}{Z_0}  \left(1 \; + \; e^{-\beta \varepsilon_{0}}
                                      \; + \: e^{-\beta(\varepsilon_{0} + E_{0})}
            \; + \; e^{-\beta(2 \varepsilon_{0} + E_{0})}\right)
\end{equation}
and $Z_0 = \sum_n exp(-E_n/k_BT) $ denotes the partition function.
The spectral weights for the bonding and antibonding
two-particle states are given by $(1-Z_f) u^2$ and
$(1-Z_f) v^2$, respectively. We stress that the form of $G_0(i\omega_n)$
 is formally equivalent to that of a BCS Green's
function, where $g^2$ plays the role of the usual gap function,
which, of course, in this lowest order approximation does not
depend on temperature.
$Z_f$  in general decreases with decreasing
temperature, which is the crucial effect that controls the
pseudogap  physics in this model. It leads, already on a
purely local electronic consideration, to a situation where
the spectral weight of the single-particle states  diminishes
with decreasing temperature, while at the same time the spectral weight
of the two-particle bonding and antibonding states  increases. These
features are reminiscent of the situation of a BCS
superconductor below $T_c$ where the role of the two-particle
states is played by the Cooper pairs.

\section{DYNAMICAL MEAN FIELD THEORY APPROACH TO THE BFM}

In developing a theoretical description which accounts for
the itinerancy of the electrons, care must be taken in fully taking
into account the local electronic structure which contains the
essential ingredients of the pseudogap phase.
This can be done by using as a starting basis the atomic limit of the
model and then introducing the effects of the
electron itinerancy within the Dynamical Mean Field Theory (DMFT)
\cite{Georges-96}. In this approach a many-body system is seen as a
purely local system coupled to a "medium",
representing a Weiss field to be determined self-consistently.
The problem is cast into the form
of a single-impurity Anderson problem described by an effective
Hamiltonian
\begin{eqnarray}
H \; = \; \sum_{\sigma} \varepsilon_{0} c_{\sigma}^{\dagger} c_{\sigma}
\; + \; E_{0} b^{\dagger} b \; + \; g \; [ \; c_{\uparrow}^{\dagger}
c_{\downarrow}^{\dagger} b \;
+ \; b^{\dagger} c_{\downarrow} c_{\uparrow}]  \nonumber \\
\; \; \; \; \; + \; \sum_{k,\sigma} w_{k}
d_{k\sigma}^{\dagger}d_{k\sigma} \; + \; \sum_{k,\sigma} \;
v_{k} \; [ \; d_{k\sigma}^{\dagger}
c_{\sigma} \; + \; c_{\sigma}^{\dagger} d_{k\sigma} \; ]
\end{eqnarray}
where $c_{\sigma}^{(\dag)}$ and $b^{(\dag)}$ denote the original
fermion and boson operators, respectively, at a single impurity site
and $d_{k\sigma}^{(\dag)}$ denote the auxiliary Fermionic operators
associated with the Weiss field. The evaluation of their energy
spectrum $w_{k}$ and their coupling $v_{k}$ to the impurity
electrons has to be performed in a self-consistent way.

It can be shown that the impurity Green's function can be cast into
the form
\begin{equation}
G_{imp} (\omega_n)= {1 \over i \omega_n - \varepsilon_0 -
\Sigma_W(\omega_n) - \Sigma_{int}(\omega_n)}
\label{gimp}
\end{equation}
which makes explicit two momentum-independent contributions to the
self-energy: one contribution being due to the boson-fermion
exchange coupling and denoted by $\Sigma_{int}(\omega)$, and one
contribution being due to the hybridization of the impurity center
with the medium and denoted by $\Sigma_W(\omega)$. This latter
quantity, depending on the parameters $v_{k}$ and $w_{k}$ and
generally referred to as the Weiss self-energy, is determined
self-consistently by equating the impurity Green's function
(\ref{gimp}) to the integral over the  ${\bf k}$ space of the
lattice Green's function

\begin{eqnarray}
G_{lat}(\omega_n, \varepsilon_{\bf k}) = {1 \over i \omega_n + \mu
-\varepsilon_{\bf k} - \Sigma_{int}(\omega_n)}
\end{eqnarray}
where $\varepsilon_{\bf k}$ denotes the bare electron dispersion.
Upon replacement of the integration over ${\bf k}$ by an integration
over energy, the integrated lattice Green's function takes the form

\begin{equation}
G_{lat}(\omega_n)= \int d\varepsilon \,{\rho(\varepsilon)
\over i\omega_n + \mu - \varepsilon - \Sigma_{int}(\omega_n)
\quad .}
\label{glat}
\end{equation}
Assuming, as is usually done, a Bethe lattice in infinite dimensions,
the density of states (DOS) for
the bare electrons appearing in eq.(\ref{glat}) is the semi-circular
DOS $\rho(\varepsilon)=(1/2\pi t^2)\sqrt{\varepsilon(4t-\varepsilon)}$
of width $2D=4t$.
The self-consistency condition $G_{imp}(\omega_n)=G_{lat}(\omega_n)$
implies
\begin{equation}
\Sigma_W(\omega_n) \; = \; t^2 G_{imp}(\omega_n)
\end{equation}
or alternatively
\begin{equation}
\Delta(\omega)\;=\;t^2\;A_F(\omega)
\label{del}
\end{equation}
where $A_F(\omega)=-2 \, {\rm Im} \, G_{imp}(\omega_n=\omega+i\delta)$
is the fermionic DOS and
\begin{equation}
\Delta(\omega) \; = \; 2 \pi \; \sum_{k} \; v_{k}^{2} \;
\delta(\omega - w_{k})
\end{equation}
is the spectral function associated with the self-energy of the
auxiliary fermions.

The above impurity problem is solved here within the
so-called Non Crossing Approximation (NCA) \cite{Bickers-87},
along the lines presented and discussed in Ref.\cite{Robin-98}.
Within the DMFT framework, NCA has recently been applied to
several other models, such as
a multiband Hubbard model for perovskites \cite{Avignon},
the Anderson lattice model with correlated conduction electrons
\cite{Schork} and the Kondo lattice model with correlated conduction
electrons \cite{Schork-99}.

One serious drawback of the DMFT+NCA procedure is that it
does not recover the non-interacting
limit, corresponding in the BFM to the case $g=0$.
The point is that this approach leads to some kinematic interactions
which result in a self-energy $\Sigma_{int}(\omega)$ which does not
vanish as the coupling constant $g$ tends to zero. In this section we
highlight this shortcoming for the fully symmetric case, realized
when the boson site energy and the chemical potential are both pinned
in the middle of the fermionic band, giving $n=2$ for all temperatures.
In the top figure of Fig.1 we plot ${\rm Im} \, \Sigma_{int}(\omega)$
at $T=0.04$ for different coupling constants $g$, as obtained within
the conventional NCA approach. We notice a strong dependence on $g$
for the frequency regime around the chemical potential, characterized
by a rapid decrease as $g$ tends to zero, as it should be. On the
contrary, for larger frequencies ${\rm Im} \, \Sigma_{int}(\omega)$
is sizeable and shows little dependence  on
the value of $g$. This fact indicates the existence of kinematical
interactions which, contrary to what one would expect, are effective
even for $g=0$ and thus should be subtracted out in the calculation
procedure. In order to do that, we first consider the case without
interaction ($g=0$) for a given temperature
and calculate self-consistently
the interaction self-energy $\Sigma^{g=0}_{int}(\omega)$ within the
standard DMFT+NCA approach. As just pointed out, this leads
to an unphysical expression which over a large range of energies
does not vanish, as it ought to in the absence of interaction.
We then repeat the whole calculation for finite boson-fermion
coupling at the same temperature, obtaining a given expression
$\Sigma_{int}(\omega)$ for the interaction self-energy. Finally, we
replace in the integrated lattice Green's function (\ref{glat}) the
self-energy $\Sigma_{int}(\omega)$ calculated at finite $g$ by
$\Sigma_{int}(\omega) - \Sigma^{g=0}_{int}(\omega)$, thus subtracting
out the unphysical part of the interaction self-energy coming from the
non-interacting case. In this way eq.(\ref{glat}) is rewritten in
the form
\begin{equation}
G^*_{lat}(\omega) = \int d\varepsilon {\rho(\varepsilon) \over
\omega + \mu - \varepsilon - \Sigma_{int}(\omega)+
\Sigma^{g=0}_{int}(\omega)} \; ,
\end{equation}
thus leading to a redefined integrated lattice Green's function which
in the non-interacting case correctly gives back the fermionic
DOS $\rho(\varepsilon)$ for free electrons.

In the bottom panel of Fig.1 we illustrate the behavior of
${\rm Im} \, \Sigma_{int}(\omega)$ as evaluated within this
redefined NCA approach. From a comparison with the curves reported
in the upper panel, we notice that as far as the low frequency
behavior around the chemical potential is concerned,
the two methods give results which are in better and better agreement
as increasing values of $g$ are considered.
Outside this frequency regime the self-energy is in general small
when evaluated within the redefined NCA, going to zero for $g=0$,
as it actually ought to be the case.


\begin{figure}
\vspace{1.5cm} \centerline{\epsfxsize=9cm \epsfbox{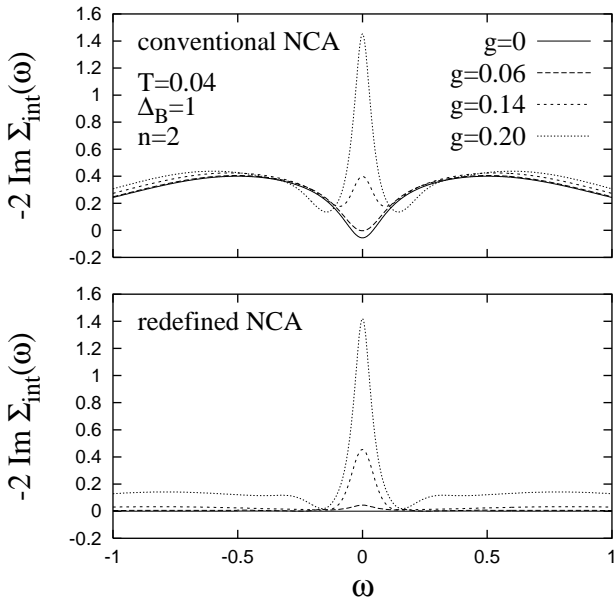}}
\caption{Comparison in the fully symmetric case $(\Delta_B=1$ and
$n=2$) of the imaginary part of the interaction
self-energy evaluated within the conventional NCA (top panel) and
the redefined NCA approach (bottom panel) for various coupling
constants $g$ and $T=0.04$}
\label{self}
\end{figure}


This difference in the self-energy also affects the behavior of the
density of states. In Fig.2 we compare the fermionic DOS evaluated
within the conventional NCA, given by
$A_F(\omega) =-2\,{\rm Im}\, G_{lat}(\omega)$, with that
evaluated within the redefined NCA approach, given by
$A^*_F(\omega)  =-2\,{\rm Im}\,
{G}^*_{lat}(\omega)$. We again notice that for small values of $g$ the
discrepancy between the two approaches is noticeable, with unphysical
tails in
the DOS obtained from the standard NCA which should not be present.
As higher values of the coupling $g$ are considered, the agreement
between the two NCA approaches is increasingly improved, in particular
in the frequency regime close to the chemical potential.


\begin{figure}
\vspace{5.5cm} \centerline{\epsfxsize=8cm \epsfbox{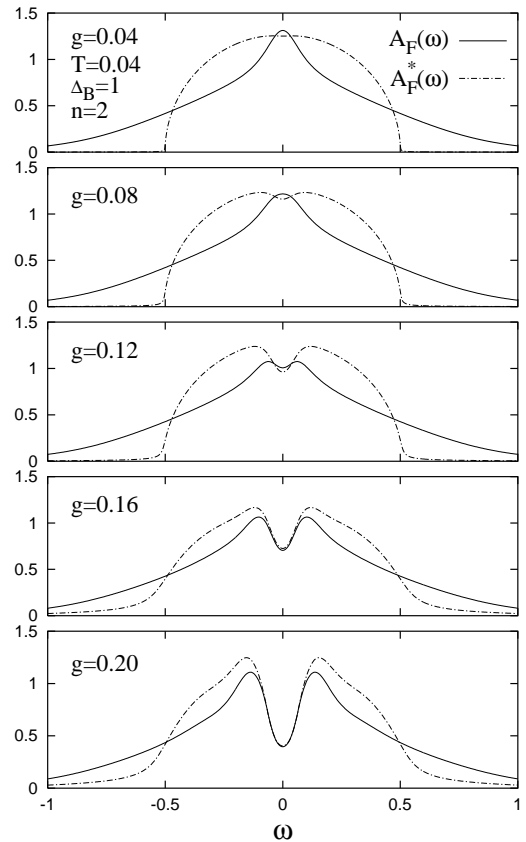}}
\caption{Comparison in the fully symmetric case $(\Delta_B=1$ and $n=2$)
of the fermionic DOS $A_F(\omega)$ and  $A^*_F(\omega)$, evaluated
respectively within the conventional and the redefined NCA approach
for various coupling constants $g$ and $T=0.04$} \label{DOS}
\end{figure}


The situation away from the fully symmetric case is numerically
more involved. We have nonetheless verified that the results obtained
in a number of specific cases for $n \neq 2$ are in agreement
with our findings for the fully symmetric case. Since here we are
interested in analyzing cases in which the hybridization constant $g$
is of the order of $0.2$, we conclude that we can safely work with
the conventional NCA approach, which in this regime is expected to
give correct results. This will be done for a variety of cases in the
next sections. In particular, our aim is to evaluate the
single-particle fermionic Green's function
from which measureable quantities can be derived, such as
the angle-resolved direct and inverse photoemission spectra
(ARPES and ARIPES, respectively),
and certain
transport coefficients for which vertex corrections, in the limit
of infinite dimensions, can be neglected.

\section{THE ARPES AND INVERSE ARPES SPECTRUM}

We examine first of all the doping and temperature dependence of
the spectral function of the lattice Green's
function for fermions  $A_F(\varepsilon_{\bf k},\omega)
= -2\,{\rm Im}\,G_{lat}(\varepsilon_{\bf k}, \omega)$.
Within the BFM scenario the doping process is
primarily controlled by a variation of the position of the
bosonic level. For a given $n$ this implies a
relative change in concentration between the bosons and fermions
which leads to a reduction of $n_F$ with decreasing $\Delta_B$.
We should stress that doping in HTS is an extremely complicated process
which involves charge transfer from the dielectric layers into the
conduction plane. Monitoring the doping process by changing the position
of the Bosonic energy level represents only a crude mechanism for it
(though the most significant one for this model). Other features
such as anisotropy effects and electronic correlations of course play
a role, but are neglected here.

The temperature and doping dependence of
the single-particle electron spectral function
$A_F(\varepsilon_k,\omega)$ is studied
for $\varepsilon_k = \varepsilon_{k_F}$, where $\varepsilon_{k_F}$ is
determined from the condition that the distribution function
$n_F(\varepsilon_k)=\int d \omega A_F(\varepsilon_k,\omega)$ is
independent on the temperature\cite{Randeria-95}. The behavior of
$A_F(\varepsilon_{k_F},\omega)$ is easily interpreted in terms of
the level spectrum of the BFM in the atomic limit\cite{Domanski-98}.
As demonstrated in Section 3, this spectrum consists
of bonding and antibonding two-particle states having energies
$E_4$ and $E_5$,
plus a non-bonding single-particle state having energy $\varepsilon_0$.
The dynamical mean field, mimicking the electron itinerancy,
broadens these levels into a continuous spectrum with peaks
approximately centered around $\varepsilon_{-}=E_4 -\varepsilon_0$,
$\varepsilon_{+}=E_5 -\varepsilon_0$  and $\varepsilon_0$ (see Fig.3).
We notice that there is a qualitative difference between the
underdoped ($\Delta_B=1.2$) and the overdoped ($\Delta_B=1$) situation
(here $\Delta_B/2$ gives the position of the energy level of the
two-particle bosonic state with respect to the bare electronic band, as
measured from its bottom).
In the case of an underdoped system  we can clearly
distinguish spectral features (Fig.3a)
with a predominant peak corresponding to the non-bonding state
which, as the temperature decreases, has its intensity transferred to
the lower-lying two-particle bonding state and thereby leads to
the opening of a pseudogap around the chemical potential. This is quite
different from what happens in the overdoped situation (Fig.3b) where
the spectral weight is roughly equally distributed between the bonding
and the antibonding two-particle states and no clear feature for the
opening of a pseudogap is visible. This
effect is particularly evident when plotting (Fig.4) the spectral function
at low temperature, together with the corresponding intensity for
Angle-Resolved Photoemission Spectroscopy
(ARPES) given by  $A_F(\varepsilon_{k_F},\omega)f(\omega)$
($f(\omega)$ denoting the Fermi function) for different
doping levels (different values of $\Delta_B$). The results in Fig.4b
are in good agreement with the experimental ARPES indication of a
pseudogap, well opened in the underdoped regime and practically not visible
in the overdoped one. In Fig.4c we plot the intensity for  Angle-Resolved
Inverse Photoemission Spectroscopy
(ARIPES) given by  $A_F(\varepsilon_{k_F},\omega)(1-f(\omega))$ which clearly
 indicates the contribution from the antibonding two-particle states above the
chemical potential. With the rapid improvement of photemission spectroscopy
experiments such a feature might be resolved in some near future.

\begin{figure}
\vspace{3.5cm} \centerline{\epsfxsize=8cm \epsfbox{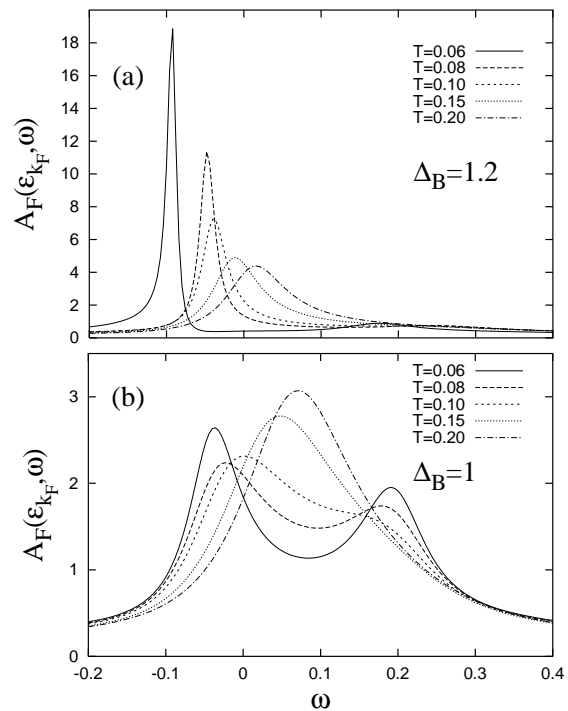}}
\caption{The electron spectral function at the Fermi surface in
the underdoped (a) and the overdoped (b) case for $g=0.2$ and $n=1.5$
(energies plotted
with respect to the chemical potential)} \label{spectralfct}
\end{figure}
Let us now turn to the discussion of the ordinary dc $(\sigma)$
and Hall $(\sigma_H)$ conductivity and show to what extent they are
influenced by the anomalous single-particle spectral properties
discussed above. In the limit of infinite
dimensionality these quantities reduce to\cite{Pruschke-95,Kotliar-98}
\begin{eqnarray}
\sigma=\frac{\pi e^2}{2 d}\int d\varepsilon\; \rho(\varepsilon) \int
d\omega\; \left( -\frac{\partial f(\omega)}{\partial \omega} \right)
A_F(\varepsilon,\omega)^2 \\
\sigma_H=\frac{\pi^2 e^3 H}{3 d^2}\int d\varepsilon\;
\varepsilon\rho(\varepsilon) \int
d\omega\; \left( -\frac{\partial f(\omega)}{\partial \omega} \right)
A_F(\varepsilon,\omega)^3 .
\end{eqnarray}
From the above expressions we evaluate, with the help of the
single-particle spectral functions $A_F(\varepsilon,\omega)$
determined before, the temperature and doping dependence of the
Hall coefficient $R_H=\sigma_H/\sigma^2 H$. In Fig.5 we plot $R_H$
as a function of temperature for different doping levels. As we
decrease $\Delta_B$ for a fixed temperature, $R_H$ decreases,
showing a sign change in the high-temperature regime below $T
\simeq 0.17$. In the underdoped regime ($\Delta_B \simeq 1.2$)
$R_H$ increases rapidly  as the temperature decreases, while a
similar temperature variation in the overdoped limit ($\Delta_B
\simeq 1$) changes $R_H$ only moderately. With decreased doping
(increasing $\Delta_B$) the temperature at which the sign change
occurs moves upwards. The sign change of $R_H$ suggests a
change-over from negative charge carriers at high temperatures to
positive charge carriers at low temperature. While at high
temperatures $R_H$ is roughly determined by the concentration of
electrons $n_F$, at low temperatures on the contrary it is roughly
controlled by $1-n_F$ and hence scales with doping.
\begin{figure}
\vspace{4.0cm} \centerline{\epsfxsize=8cm \epsfbox{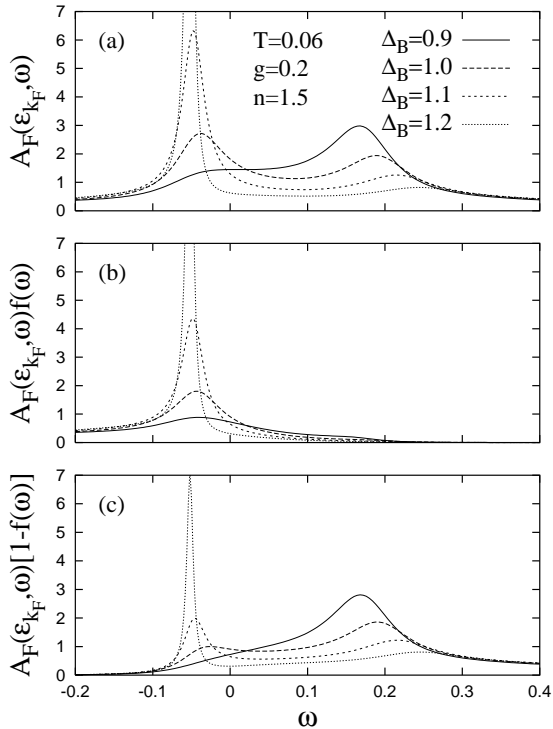}}
\caption{The low-temperature electron spectral function at the
Fermi surface (a) and the corresponding ARPES (b) and ARIPES (c)
intensities for $n=1.5$, $g=0.2$ and different doping levels
(different $\Delta_B$). The energies are measured relative to the
chemical potential.} \label{spectralfct1}
\end{figure}
Although experiments seem to suggest
that for underdoped systems $R_H$ is always positive, our results
indicate that as doping is reduced the sign change occurs for higher
and higher temperatures, possibly not reachable in real experiments.
At first sight it is tempting to interprete the results of Fig.5
as a change-over from a large
Fermi surface at high temperatures to small Fermi surface pockets
at low temperatures. However, our study of the single-particle
spectral function indicates that the Fermi surface is always large,
in agreement with experimental findings.

We finally address ourselves to the temperature dependence of the
Hall angle $cot \, \theta_H$, which in the  underdoped regime
shows a $T^2$ behavior for a large variety of HTS, sometimes
extending up to 400 K. In Fig.6 we plot our calculated Hall angle
as a function of temperature for three different doping levels. We
notice that deep inside the underdoped regime we do indeed find a
$T^2$ behavior while with increased doping, going from
$\Delta_B=1.26$ to $\Delta_B=1.18$, the range of this $T^2$
behavior gets more and more restricted to low temperatures, in
qualitative agreement with the experiments.
\begin{figure}
\centerline{\epsfxsize=7cm \epsfbox{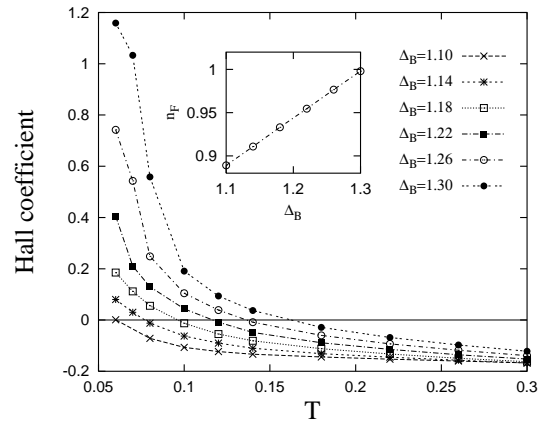}}
\caption{Evolution of the Hall coefficient as a function of
temperature for $n=1.5$, $g=0.2$ and different doping levels. The
inset shows the variation of $n_F$ with $\Delta_B$ for $T=0.26$.}
\label{R_H}
\end{figure}

\begin{figure}
\centerline{\epsfxsize=7cm \epsfbox{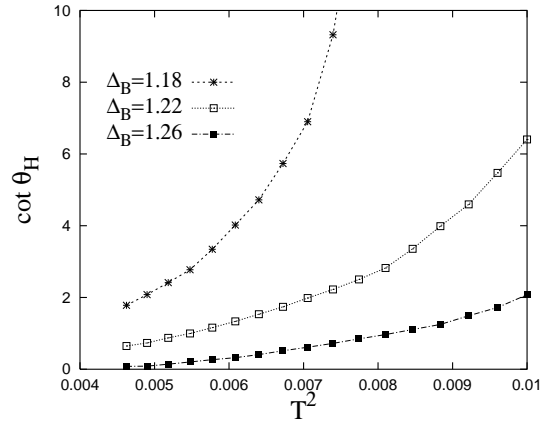}}
\caption{Evolution of the Hall angle as a function of temperature
for $n=1.5$, $g=0.2$ and different doping levels.} \label{cot_H}
\end{figure}

\section{CONCLUSIONS}

The non-Fermi liquid properties of the cuprate HTS seem now to be
established. Under these circumstances the evaluation of transport
coefficients, such as the ones discussed here, can no longer be
based on the standard Boltzmann-type relaxation time approach. The
transport coefficients are no longer determined exclusively by
scattering processes in the immediate vicinity of the Fermi
surface, as is the case for standard Fermi liquids. On the
contrary, scattering processes cover a wide regime in frequency
and wave vector, in accordance with the experimental evidence for
very broadened single-particle spectral functions. In this paper
we have examined the question to what extent the anomalous
properties of the Hall coefficient $R_H$ can be linked to the
anomalous single-particle properties and in particular to the
pseudogap features of them. Our main result is a strongly
temperature dependent $R_H$ which, for low temperatures, is
positive (hole-like charge carriers) and scales with the number of
holes as measured from the half-filled band case. As the
temperature is increased, $R_H$ rapidly drops and saturates at
very small negative values for temperatures above that where the
pseudogap opens. Correspondingly, Fermi liquid properties begin to
be recovered. Since we do not take into account electronic
correlations, this result is what one should expect. Taking
account of correlations should alter this result as far as the
doping dependence of $R_H$ is concerned and should make $R_H$
scale with the number of holes (measured from the half-filled band
case) even for high temperatures. This, however, requires a
treatment of correlations and pair fluctuations on the same
footing, which is beyond the present study. Finally, a $T^2$
behavior of the Hall angle is obtained, but exclusively for very
low doped systems.

\section{ACKNOWLEDGEMENTS}

We would like to thank T. Domanski and K. Matho for stimulating discussions.

\end{multicols}


\begin{references}
\bibitem{experiments} H. Takagi, T. Ido, S. Ishibashi, M. Uota, S. Uchida,
and Y. Tokura, Phys. Rev. B {\bf 40}, 2254 (1989); T. Nishikawa,
J. Takeda, and M. Sato, J. Phys. Soc. Japan {\bf 62}, 2568 (1993);
A. Carrington, D.J.C. Walker, A.P. Mackenzie, and J.R. Cooper,
Phys. Rev. B {\bf 48}, 13051 (1993); H.Y. Hwang, B. Batlogg, H.
Takagi, H.L. Kao, J. Kwo, R.J. Cava, J.J. Krajewski, and W.F. Peck
Jr., Phys. Rev. Lett. {\bf 72}, 2636 (1994).
\bibitem{Anderson-91} P.W. Anderson, Phys. Rev. Lett. {\bf 67},
2092 (1991); N.H. Hussey, J.R. Cooper, J.M. Wheatley, I.R. Fisher,
A. Carrington, A.P. Mackenzie, C.T. Lin, and O. Milat, Phys. Rev.
Lett. {\bf 76}, 122 (1996).
\bibitem{Pruschke-95} T. Pruschke, M. Jarrell and J. Freericks,
Adv. Phys. {\bf 44} 187 (1995).
\bibitem{Kotliar-98} E. Lange and G. Kotliar, Phys. Rev. B {\bf 59},
1800 (1999).
\bibitem{Shastry-93} B.S. Shastry, B.I. Shraiman and R.R.P. Singh,
Phys. Rev. Lett. {\bf 70}, 2004 (1993).
\bibitem{Corson-98} J. Corson, R. Mallozzi, J.N. Eckstein, I. Bozovic,
and J. Orenstein, Nature {\bf 398}, 221 (1999).
\bibitem{Randeria-98} M. Randeria, Proc. Int. School of Physics
"Enrico Fermi", Course CXXXVI, eds. G. Iadonisi, R. Schrieffer and
M.L. Chiofalo (IOS Press, Amsterdam, 1998) pp. 53-75; O.
Tchernyshyov, Phys. Rev. B {\bf 56}, 3372 (1997); R. Micnas, M.H.
Pedersen, S. Schafroth, T. Schneider, J.J. Rodr\'iguez
N\'u$\tilde{\rm n}$ez, and H. Beck, Phys. Rev. B {\bf 52}, 16223
(1995); B. Janko, J. Mali and K. Levin, Phys. Rev. B {\bf 56},
R11407 (1997); M. Letz and R. J. Gooding, J. Phys.: Condens.
Matter {\bf 10}, 6931 (1998).
\bibitem{Ranninger-85} J. Ranninger and S. Robaszkiewicz, Physica B
{\bf 135}, 468 (1998).
\bibitem{Ranninger-95}  J. Ranninger, J.-M. Robin and M. Eschrig,
Phys. Rev. Lett. {\bf 74}, 4027 (1995), J. Ranninger and J.-M.
Robin, Solid State Commun. {\bf 98}, 559 (1996), J. Ranninger and
J.-M. Robin, Phys. Rev. B {\bf 53} R11961 (1996).
\bibitem{Domanski-98} T. Domanski, J. Ranninger and J.-M. Robin,
Solid State Commun. {\bf 105}, 473 (1998).
\bibitem{Georges-96} see for instance: A. Georges, G. Kotliar,
W. Krauth and M. Rozenberg, Rev. Mod. Phys. {\bf 68}, 13 (1996);
F. Gebhard, "The Mott Metal-Insulator Transition" (Springer Verlag,
berlin, 1997).
\bibitem{Bickers-87} N.E. Bickers, Rev. Mod. Phys. {\bf 59}, 845 (1987).
\bibitem{Robin-98} J.-M. Robin, A. Romano and J. Ranninger,
Phys. Rev. Lett. {\bf 81}, 2756 (1998).
\bibitem{Avignon} P. Lombardo, J. Schmalian, M. Avignon, and K.-H.
Bennemann, Phys. Rev. B {\bf 54}, 5317 (1997).
\bibitem{Schork} T. Schork and S. Blawid, Phys. Rev. B {\bf 56},
6559 (1997).
\bibitem{Schork-99} T. Schork, S. Blawid, and J. Igarashi,
Phys. Rev. B {\bf 59}, 9888 (1999).

\bibitem{Randeria-95} M. Randeria, G. Jennings, H. Ding, J.-C. Campuzano,
A. Bellman, T. Yokoya, T. Takahashi, H. Katayama-Yoshida, T.
Mochiku, and K. Kadowaki, Phys. Rev. Lett. {\bf 74}, 4951 (1995).
\bibitem{Ino-98} A. Ino, C. Kim, T. Mizokawa, Z.-X. Shen, A Fujimori,
M. Takaba, K. Tamasaku, H. Eisaki, and S. Uchida, J. Phys. Soc.
Japan {\bf 68}, 1496 (1999).
\bibitem{Hallangle} T. Chien, Z.Z. Wang, and N.P. Ong, Phys. Rev. Lett.
{\bf 67}, 2088 (1991); G. Xiao, P. Xiong, and M.Z. Cieplak, Phys.
Rev. B {\bf 46}, 8687 (1992); B. Bucher, P. Steiner, J. Karpinski,
E. Kaldis, and P. Wachter, Phys. Rev. Lett. {\bf 70}, 2012 (1993);
J.M. Harris, H. Wu, N.P. Ong, R.L. Meng, and C.W. Chu, Phys. Rev.
B {\bf 50}, 3246 (1994).

\end{references}
\end{document}